\def\LCDM{$\Lambda$CDM}
\def\lesssim{\mathrel{\hbox{\rlap{\hbox{\lower4pt\hbox{$\sim$}}}\hbox{$<$}}}}
\def\gtrsim{\mathrel{\hbox{\rlap{\hbox{\lower4pt\hbox{$\sim$}}}\hbox{$>$}}}}

\newcommand{\AAA}[3]    {#3, A\&A, \textbf{#1}, #2}

\newcommand{\AAS}[3]    {#3, A\&A~Suppl., \textbf{#1}, #2}
\newcommand{\ApJ}[3]    {#3, ApJ, \textbf{#1}, #2}
\newcommand{\ApJS}[3]   {#3, ApJ~Suppl., \textbf{#1}, #2}

\newcommand{\AJ}[3]     {#3, Astron.~J., \textbf{#1}, #2}
\newcommand{\MNRAS}[3]  {#3, MNRAS, \textbf{#1}, #2}
\newcommand{\Nature}[3] {#3, Nature, \textbf{#1}, #2}

\newcommand{\astroph}[1]{\texttt{astro-ph/#1}}

\documentstyle[epsfig]{mn}

\begin{document}

\title{Collision-induced galaxy formation:  semi-analytical model and multi-wavelength predictions}

\author[Balland C. et al.]
       {Christophe Balland$^{1}$, Julien E. G. Devriendt$^{2}$,
        and Joseph Silk$^{2}$\\
       {$^1$Universit\'e Paris Sud, IAS-CNRS, B\^atiment 121,
        91405 Orsay Cedex, France}\\
       {$^2$University of Oxford, Astrophysics, Keble Road, Oxford, OX1 3RH, UK}}

\date{Received ...; accepted ...}

\maketitle

\begin{abstract}
A  semi-analytic model is proposed that couples  the Press-Schechter
formalism for the number of galaxies with a prescription for
galaxy-galaxy interactions that enables to follow the evolution of galaxy
morphologies along the Hubble sequence.  Within this framework, we
calculate the chemo-spectrophotometric evolution of galaxies to obtain
spectral energy distributions. We find that such an approach is very
successful in reproducing the statistical properties of galaxies as well
as their time evolution. We are able to make
predictions as a function of galaxy type: for clarity, we restrict ourselves
to two categories of galaxies:
early and late types  that are  identified with ellipticals and disks.
In our model, irregulars are simply an early stage  of galaxy formation.
In particular, we obtain good matches for
the galaxy counts and  redshift distributions of sources
 from UV to submm wavelengths.  We  also
reproduce  the observed cosmic star formation history
and the diffuse
background radiation,
and make predictions as to the epoch and wavelength at which
 the dust-shrouded star formation of
spheroids begins to dominate over the star formation
that occurs more
quiescently in disks.
A new prediction  of our model is a rise in the FIR luminosity
density with
increasing redshift, peaking at about $z\sim 3$,
and with
a ratio to the local luminosity density $\rho_{L,\nu} (z = z_{peak})/ \rho_{L,\nu} (z = 0)$
  about 10 times higher than that in the blue (B-band)
which peaks near $z\sim 2$.
\end{abstract}


\section{Introduction}

In contrast with the local universe, where only 30 \% of the
bolometric luminosity is released in the IR/submm wavelength range
(Soifer \& Neugebauer 1991), there is a growing amount of evidence
that the high--redshift universe was much more opaque. Indeed, the
discovery of the Cosmic Infrared Background (CIRB) at a level ten
times higher than the no--evolution predictions based on the {\sc
iras} local IR luminosity function, and twice as high as the Cosmic
Optical Background obtained from optical counts, has shown that dust
extinction and emission are key processes for high--redshift galaxies
(Puget et al. 1996; Guiderdoni et al. 1997;
Fixsen et al. 1998; Hauser et al. 1998; Schlegel, Finkbeiner
\& Davis 1998). Deep surveys with the {\sc
iso} satellite at 15 $\mu$m (Oliver et al. 1997; Aussel et al. 1999;
Elbaz et al. 1999) and 175 $\mu$m (Kawara et al. 1998; Puget et al.
1999), and with the SCUBA instrument at 850 $\mu$m (Smail, Ivison \& Blain
1997; Barger et al. 1998; Hughes et al. 1998; Eales et al. 1999;
Barger, Cowie \& Saunders 1999) have begun to resolve  the CIRB into its brightest
contributors. Although identification and spectroscopic follow--up of
submm sources are not easy, such studies seem to reach the conclusion
that an important fraction of these sources are the high--redshift
counterparts of the local luminous and ultraluminous IR galaxies
(LIRGs and ULIRGs) discovered by {\sc iras} (Smail et al. 1998; Lilly
et al. 1999; Barger et al. 1999).  In the optical and near--IR
window, careful examinations of the Canada--France Redshift Survey
(CFRS) galaxies at $z\sim 1$, and Lyman break galaxies at $z \sim 3$
and 4 have revealed a significant amount of extinction (Flores et al.
1999; Steidel et al. 1999; Meurer, Heckman \& Calzetti 1999).
This has lead to a
reassessment of previous estimates of the UV fluxes, and consequently
of the star formation rates in these objects, which are now found to
be higher by a factor 2 to 5.  In light of these observations, one can
view the far infrared background as a sink for the hidden aspects of
galaxy formation.  At optical wavelengths, ellipticals and spheroids
are old, even at $z \sim 1.$  No evidence is seen for either
 the luminous formation phase, or
the early evolution at these wavelengths. One  therefore
draws the conclusion that ellipticals and, more generally, most
spheroids must have formed in dust-shrouded starbursts.

In this paper, we show how separate tracking of disk and spheroid star
formation within a simple but explicit cosmological framework enables
one to substantiate this view and show that current data is consistent
with spheroid formation dominating the submillimetre background.  In
section 2, we briefly describe how we connect the various pieces of
our semi-analytic model together, to differentiate between spheroids
and disk galaxies and use the {\sc stardust} model galaxy spectra
(Devriendt, Guiderdoni \& Sadat 1999) to make the link with
observations.  The approach is very similar in spirit to that of
Devriendt \& Guiderdoni (2000) - DG00 hereafter, so we point out the
differences/improvements with respect to this previous work.  Section
3 presents the results obtained with this new implementation of
collision-induced galaxy formation.
Finally, we put our work in context  with
other studies and
draw conclusions in Section 4.

\section{Overview of the model} 

\subsection{Cosmological framework}

To match the combined observations of the cosmic microwave background
anisotropies on sub-degree scales by BOOMERanG (de Bernardis et
al. 2000) and MAXIMA (Balbi et al. 2000), and of high-redshift
supernovae (Riess et al. 1998; Schmidt et al. 1998; Perlmutter et
al. 1999), we have chosen a flat universe model with a cosmological
constant. More specifically, the ratio of total matter density to
critical density and the reduced cosmological constant are taken to be 
$\Omega_0 = 0.3$ and $\lambda_0 = 0.7$ respectively. 
The Hubble constant $H_0$ was chosen to lie within
the error bar quoted by the Hubble Space Telescope key project value
(Freedman et al. 2001). 
The value of $\Omega_B = 0.02 \, h^{-2}$ is the one currently favored
by various cosmological probes of Big Bang
nucleosynthesis (Bania et al. 2002). Finally the normalization of our non-tilted primordial 
Harrison-Zeldovich power spectrum $\sigma_8 = 0.86$ ensures fair agreement both with the 
amplitude of the cosmic background radiation measured by COBE and the local cluster
abundance. We note that recent cluster surveys seem to favor
slightly lower values around $\sigma_8 \sim 0.73$ (e.g. Lahav et al. 2002) but
our conclusions are quite insensitive to a change of 20 \% in 
any of the parameters presented in table~\ref{cosmoparam}.

\begin{table}
\caption{Parameters of the cosmological model. $h$ is the
Hubble constant at present in units of 100 km/s/Mpc. $\sigma_8$
is the rms mass fluctuation in spheres of 8$h^{-1}$ Mpc.}
\label{cosmoparam}
\begin{tabular}{lllll} \hline \hline
$\Omega_0$   & $\lambda_0$ & $h$ & $\sigma_8$ & $\Omega_B$   \\ \hline
 0.3  & 0.7 & 0.65& 0.86 & 0.02 $h^{-2}$     \\ \hline \hline
\end{tabular}
\end{table}

In the semi-analytical  {\em ab initio} approach that we outline in the next
section, galaxies form from Gaussian random density fluctuations in
the primordial matter distribution, dominated by cold dark matter
(CDM).  To follow a bound perturbation, we use the well known
spherical top-hat collapse model: a homogeneous spherical
perturbation grows along with the expanding universe, until
self--gravity results in  turn around and (non--dissipative)
collapse. We then compute the number of perturbations of a given mass
which collapse at a given redshift using the standard Press \& Schechter
(1974) prescription and starting with a primordial power spectrum for
the density field given by Bardeen et al. (1986), corrected for the
presence of baryons (Sugiyama 1995).  To take into account mass build up 
of halos through mergers, we employ the extended Press--Schechter formalism 
described in e.g. Lacey \& Cole (1993), in a similar fashion as  
in Balland, Silk \& Schaeffer (1998) - BSS98 hereafter.
Note that this is slightly
different from Devriendt \& Guiderdoni (2000), where the
prescription to compute the number of collapsed haloes was given by
the peaks formalism developed in Bardeen et al. (1986) and where
halo merging histories were not considered.

\subsection{A simple semi-analytic model}

To make contact with observations, star formation and spectral
evolution must be incorporated. We follow the general framework
described by White \& Frenk (1991), that is to say, within virialized
dark matter halos, we let gas cool radiatively, settle into a disk and
form stars. Technical details about the semi-empirical recipes used to
model the different astrophysical processes are given in
DG00. Chemical and spectral evolution  are then computed with the
{\sc stardust} stellar population synthesis model from Devriendt et al. (1999).
Detailed references can be looked up in this paper, but, as our
results in the far-IR/submm crucially depend upon it, we summarize 
the basic assumptions/ingredients concerning dust that go into this modelling.
The first of these is that metals and dust are homogeneously mixed with stars
in galaxies of oblate ellipsoidal shape. The extinction curve of each galaxy 
is then assumed to scale with its gas and metal contents, in a manner that is calibrated 
on the Milky Way and nearby spirals. From such a model, the radiative transfer 
can be computed analytically, and yields the total amount of stellar light absorbed at 
UV/optical wavelength. 
Following energy conservation, this energy must be re-emitted at IR wavelengths.
Implementing the 3-component absorption/emission model of D\'esert, Boulanger \& Puget (1990),
(a mixture of PAHs, very small grains and large grains) as in {\sc stardust}, i.e.  in 
combination with the empirical correlation of the bolometric IR luminosity with {\sc iras}
colours, we finally link the optical and the far-IR/submm windows in a self--consistent way.

We emphasize that, as discussed in detail by DG00, there are three key
parameters in such models, the star formation efficiency $\beta^{-1}$,
the feedback efficiency $\epsilon_{SN}$ and the extent of the gaseous
disks $f_c$. We take parameter values that are fairly similar to the
ones used by these authors, with $\beta = 60$, $\epsilon_{SN} = 0.2$
and $f_c = 5$.  These values are well within the uncertainties of
observations by Kennicutt (1998) and Bosma (1981) for the star
formation efficiency and the gaseous extent of cold disks,
respectively. Numerical simulations by e.g. Thornton et al. (1998)
tend to give values closer to 0.1 for feedback efficiency; however our
higher value is a consequence of an attempt to reduce the number of
small objects overestimated by the Press-Schechter prescription, as
well as a means to cure the ``cooling catastrophe'', where too much
gas cools in low mass halos at high redshift.  The qualitative effect
of each of these parameters can be summed up in the following way:
\begin{itemize}
\item  Increasing $\beta$ decreases the normalisation and faint-end slope of
the optical and IR counts, because star formation is reduced and takes
place at lower redshift.
\item Increasing $\epsilon_{SN}$ decreases the normalisation and faint-end slope of the
optical and IR counts by quenching star formation in higher and higher
mass galaxies that form on average at lower and lower redshifts.
\item Increasing $f_c$ means increasing the
normalisation of the optical counts, and decreasing  that of the IR
counts, since this is equivalent to reducing extinction.
\end{itemize}

The main drawback of DG00 is their modelling of ultraluminous dusty
starbursts: they use an {\em ad hoc} fraction of massive objects
$F$, calibrated locally on the {\sc iras} luminosity function and which is
evolving with redshift as $F \propto (1+z)^6$.  In this paper, we
replace this ``recipe'' by a physically motivated collision model which
allows us to differentiate between disks and ellipticals.

\subsection{Modelling galaxy types}

To separate disk galaxies from spheroids we use the prescription
proposed in BSS98.
It is based on the amount of energy exchanged by a galaxy
with its neighbours via gravitational interactions during
its lifetime. Interactions are modelled as rapid non-merging
collisions following the Spitzer impulse approximation (Spitzer 1958).
If we denote by
${\dot \Delta}$ the instantaneous relative rate of energy exchange,
$(1/E)dE/dt$, at epoch $t$, the cumulative effect
of interactions occurring throughout  the galaxy lifetime  (from its birth
at the redshift $z_{nl}$ at which its host halo collapses to the redshift $z$ of observation) is obtained
by integration:

\begin{equation}
\label{eqdelt}
\Delta\equiv \int \frac{dE}{E}=\int_{z_{nl}}^z{\dot \Delta}\frac{dt}{dz}dz
\end{equation}

In the above expression, the main parameters describing an interaction
(impact parameter, relative velocity between the interacting
galaxies, relative mass, number density of neighbouring
galaxies) enter through the quantity ${\dot \Delta}$.
The scalings with these parameters are derived from the
numerical simulations of Aguilar \& White (1985). In this way, it
is possible to overcome the difficulty of modelling analytically the details
of the interaction, which is a highly non-linear process.

The effects of cosmology enter equation (\ref{eqdelt}) via the
conversion of the loockback time $t$ into redshift $z$ and the scaling
of the interaction parameters with redshift. For instance, in the
present cosmology (see table 1), the relative velocity between two
interacting galaxies is given by:

\begin{equation}
\label{eqvrel}
v=v_0 \sqrt{\Omega_0(1+z)+\lambda_0(1+z)^{-2}}.
\end{equation}

The evaluation of equation (\ref{eqdelt}) for a galaxy in the
field (see BSS98 for details) gives (in a flat, $\Omega_0+\lambda_0=1$,
universe):

$$
\Delta(z)=\Delta_\star \Omega_0\Big\{
\frac{(1+z_{nl})^3-(1+z)^3}{\Omega_0^2}
$$

$$
-\frac{\lambda_0^2}{\Omega_0^2}
\Big [\frac{(1+z)^3-(1+z_{nl})^3}{(\Omega_0(1+z_{nl})^3+\lambda_0)
(\Omega_0(1+z)^3+\lambda_0)}\Big ]
$$

\begin{equation}
\label{eqdelt2}
-2\frac{\lambda_0}{\Omega_0^3}\ln \Big [\frac{\Omega_0(1+z_{nl})^3+\lambda_0}
{\Omega_0(1+z)^3+\lambda_0}\Big ]
\Big\}
\end{equation}
with $\Delta_\star\approx 1.35\times 10^{-4}$ for $h=0.65$.
It is easy to verify that in the limit $\lambda_0 \rightarrow 0$
equation (\ref{eqdelt2}) reduces to $\Delta(z) = \Delta_\star[(1+z_{nl})^3-
(1+z)^3]$ which is the result for an $\Omega_0=1$ universe
(BSS98).

Morphological types are then defined according to the
value of $\Delta$ using the following rules:
\begin{itemize}
\item Galaxies for which $\Delta$ is lower than a certain threshold
value $\Delta_{spi}$ are identified as disks: they correspond to
galaxies having experienced few, if any, interactions and consequently
have a small $\Delta$.
\item Galaxies for which $\Delta$ is higher than a certain threshold
value $\Delta_{ell}>\Delta_{spi}$ are identified as ellipticals: they
correspond to galaxies having undergone substantial energy exchange
during their lifetime.
\item Galaxies such that $\Delta_{spi}<\Delta(z)<\Delta_{ell}$ are
identified as S0 galaxies.
\end{itemize}

The values of $\Delta_{spi}$ and $\Delta_{ell}$ are fixed by requiring
that the model produces the observed amount of each type in the field
at the present epoch. Typical values are $\Delta_{spi}\approx 3.10^{-3}$
and $\Delta_{ell}\approx 10^{-2}$, and are not very
sensitive to the cosmological model assumed (standard or \LCDM).
The above conditions on $\Delta$ can then be inverted to obtain
conditions on the formation redshift $z_{nl}(z)$ of galaxy types.  It
is important to realize that the formation redshift of a galaxy of a
given type at epoch $z$ depends on $z$. That is to say that in our
model the morphology of a galaxy is not fixed once and for all by initial
conditions but, on the contrary, it evolves progressively from
late-type to early-type. However, contrary to most other semi-analytic
approaches, this progression is monotonic: once a spiral galaxy has become  
SO it can only evolve to the elliptical stage, and it can only remain elliptical 
once it reaches that stage.   
The condition for a galaxy to be elliptical at redshift $z$ is that its formation redshift $z_{nl}$
is such that:
\begin{equation}
\label{eqzell}
z_{nl}>z_{ell}(z)
\end{equation}
where $z_{ell}(z)$ is defined by equating $\Delta(z)$ to $\Delta_{ell}$
and solving for $z_{nl}$.
Equivalently, spirals at epoch $z$ are those galaxies that turned
non-linear at an epoch such that:
\begin{equation}
\label{eqzspi}
z_{nl}<z_{spi}(z)
\end{equation}
where $z_{spi}(z)$ is defined by equating $\Delta(z)$ to $\Delta_{spi}$.
Finally, S0 galaxies are such that:
\begin{equation}
\label{eqzs0s}
z_{spi}(z)<z_{nl}<z_{ell}(z).
\end{equation}

\begin{figure}
\centerline{\psfig{file=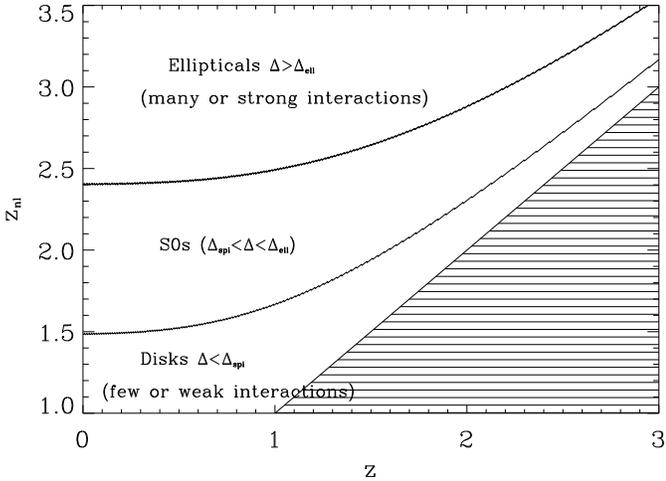,width=1.2\hsize}}
   \caption{($z_{nl},z$) plane. The upper solid curve is for
$\Delta=\Delta_{ell}$ and the lower one for $\Delta=\Delta_{spi}$.
The upper region is the locus for galaxies having
experienced high energy exchanges due
to gravitational interactions since they turned non-linear
(the ellipticals region according
to the present model), the lower region is the locus for
low energy exchange (the disks region). In between, one finds
the region for S0s. The hatched area is the region for which
$z>z_{nl}$ and is therefore not populated by galaxies.}  \label{figcutsz}
\end{figure}

Figure~\ref{figcutsz} illustrates the way morphological types are
defined in this model. The formation redshift
is shown as a function of the epoch $z$ considered. The solid curves are obtained for
$\Delta(z)=\Delta_{ell}$ (upper) and $\Delta(z)=\Delta_{spi}$ (lower).
Ellipticals occupy the upper region of the ($z_{nl},z$) plane,
while disks occupy the lower region. S0s are located
in between. The hatched area corresponds to $z_{nl}<z$ and is
irrelevant.

Consider the example of a galaxy that became non-linear
at, say, $z_{nl}=3$. Figure~\ref{figcutsz}  shows that up to
redshift $z\sim 2.8$ this galaxy is identified as a disk.
Between $z\sim 2.2$ and $z\sim 2.8$, the same galaxy has
experienced sufficient energy exchange with its neighbours
to be identified as an S0 galaxy, but not quite enough to
be an elliptical. By $z\sim 2.2$, it has become an elliptical.

Depending on the previously identified morphological type, we decide
which galaxies will undergo an ``obscure starburst''. In other words,
this phase is triggered when the energy imparted by encounters/collisions 
is strong enough to alter the morphology for the first time, i.e. when late-type 
(spiral) galaxies first become early type (S0s). The transition from 
SO to elliptical is then supposed to happen smoothly, without another burst 
of star formation, as the gas poor SO galaxies are assumed to have reheated their surroundings 
enough during the burst to not be able to replenish their gas supplies through accretion of new material.
The intensity and duration of this LIRG/ULIRG phases, 
are therefore controlled by the amount of gas available for star formation
and the size of the galaxy at the time when the morphology change occurs.  
In practice, this dark phase is
modelled by setting our three key parameters to $\beta = 1$, $f_c = 1$
and $\epsilon_{SN} = 0.5$, corresponding to high star formation
efficiency, high dust opacity, and high feedback efficiency
respectively. It is therefore completely coupled to the starburst, and as the luminosity 
of this latter decreases for lack of fuel (gas) supply, the optical depth of the galaxy 
goes down as well, revealing more and more of the stellar population at optical wavelengths.

\section{Results} 

\subsection{Galaxy counts} 

\begin{figure*}
   \centerline{\psfig{file=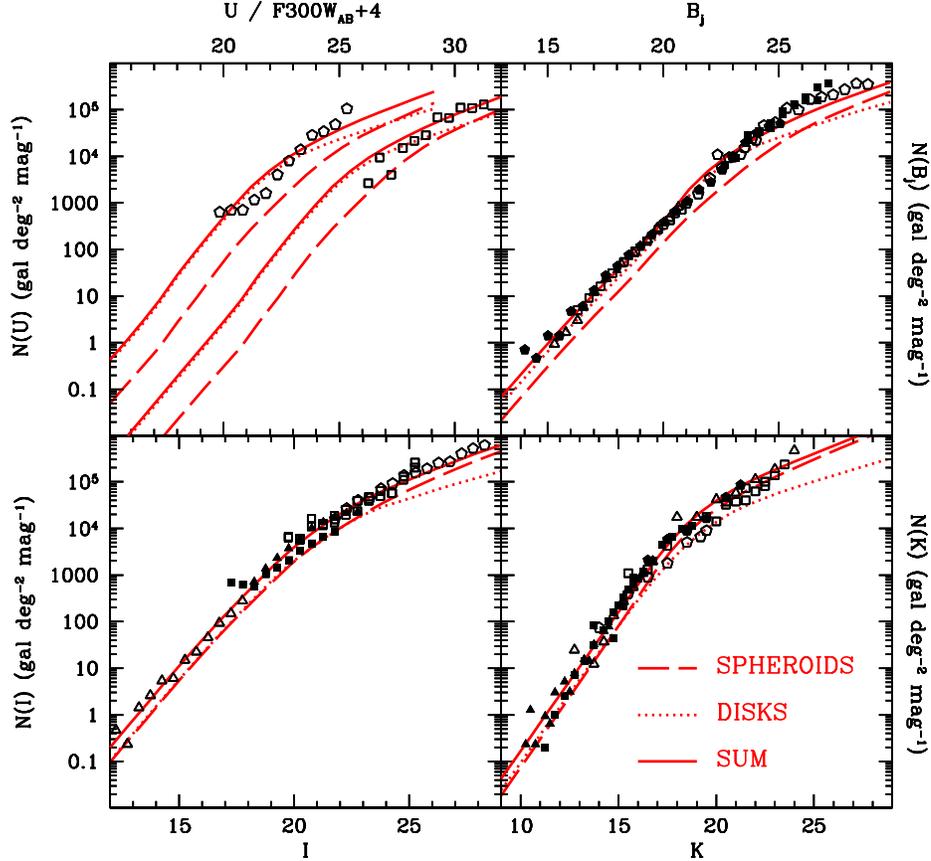,width=0.7\hsize}}
   \caption{UV/Optical/Near-IR counts from our analytic model,
   compared to existing data.  The dotted curves stand for late type
   galaxies (disks); the dashed ones for early type galaxies
   (spheroids); and the solid ones for the sum of both contributions.
   Data is from Hogg et al. (1997) (U band), Williams et al. (1996)
   (F300W$_{\rm AB}$, B \& I bands), Arnouts et al. (1997) (B band),
   Bertin \& Dennefeld (1997) (B band), Gardner et al. (1996) (B, I \&
   K bands), Metcalfe et al. (1996) (B band), Weir et al. (1995) (B
   band), Smail et al. (1995) (I band), Crampton et al. (1995) (I
   band), Moustakas et al. (1997) (K band), and Djogorvski et
   al. (1995) (K band).}  \label{figopt}
\end{figure*}

\begin{figure*}
   \centerline{\psfig{file=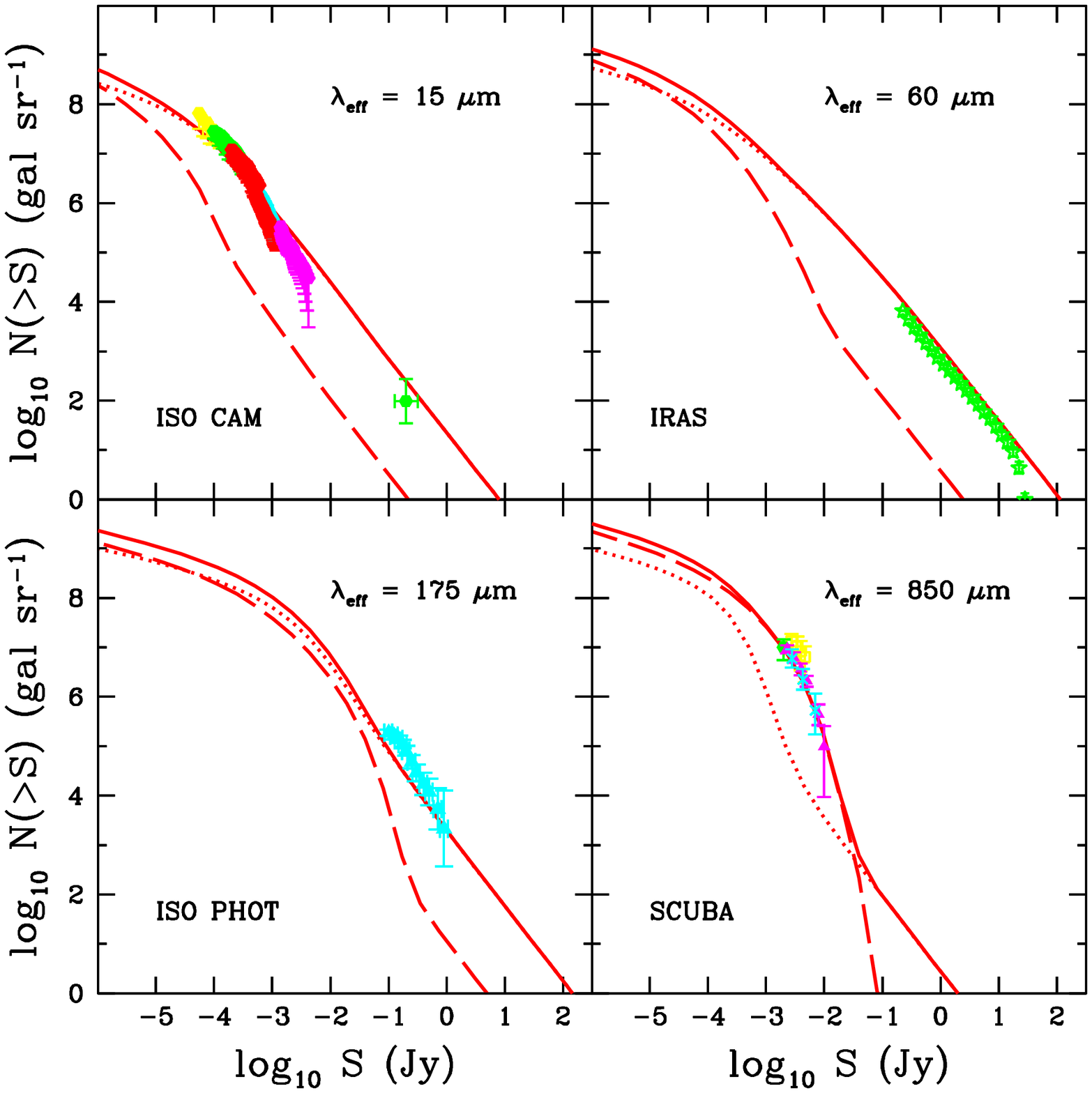,width=0.7\hsize}}
   \caption{Infrared/submm counts from the analytic model described in
   the text.  Coding for the lines is dotted: late type galaxies
   (spirals); dashed: early type (ellipticals, S0s); and
   solid: sum of both contributions.  Data are from Elbaz et
   al. (1999) (15 $\mu$m), Kawara et al. (1998), Puget et
   al. (1999) and Dole et al.(2001) (175 $\mu$m), Smail et al. (1997), Eales et al. (1999),
   Barger et al. (1999), Scott et al. (2002) (850$\mu$m).}  \label{figinf}
\end{figure*}

Looking at figures~\ref{figopt} and \ref{figinf}, one realises that
(except in the near IR bands), late type galaxies
dominate over early types (compare the dotted curves with the dashed
ones in each panel). This domination extends down to the far-IR, with
the late-type galaxies still dominating the 175 micron ISOPHOT
counts.  At longer wavelengths, however, there is a dramatic change:
the early-type galaxies completely swamp the contribution from late
types. Indeed, one can see on the bottom right panel of figure~\ref{figinf} that
the vast majority ($\approx$ 90 \%) of the SCUBA sources are
classified as early types in our model.  The reason for such a change
of behaviour lies in the well known negative k-correction, which makes
galaxies of the same bolometric luminosity  as bright for the
observer at redshift 5 as at redshift 0.5.  This is only important in
the submm (here for SCUBA at 850 microns), because the peak emissivity
of dust in the source rest frame is between 60 and 100
microns. Therefore, as our S0s/ellipticals approximately
form at $z > 2$, the corresponding
maxima of emission must be redshifted to wavelengths greater than 180
and 300 microns respectively.

This result is quite robust in the sense that its qualitative features
do not depend on the cosmological parameters. However, quantitatively,
there is a marked difference: the domination of late type galaxies is
more marked in a SCDM model, where galaxies tend to form later on
average (see Silk \& Devriendt 2000). This remark also applies to
the model of DG00, where the phenomenologically evolving 
ULIRG fraction was the dominant contributor in the far-IR (175 microns).

\begin{figure*}
   \centerline{\psfig{file=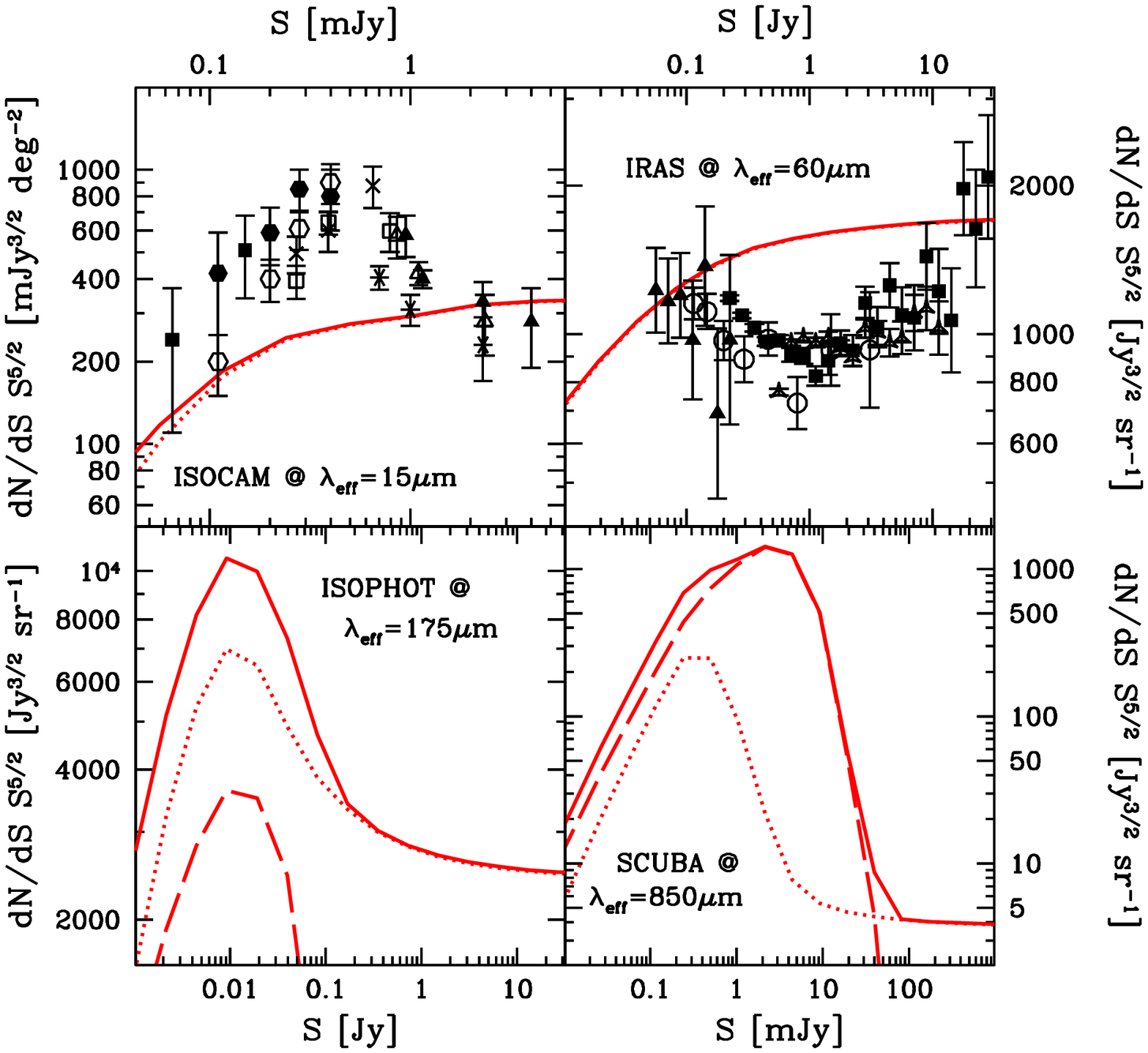,width=0.6\hsize}}
   \caption{Infrared/submm differential counts
   from the model.
   Coding for the lines is dotted: late type galaxies
   (spirals); dashed: early type (ellipticals, S0s); and
   solid: sum of both contributions.  Data are taken from various
   {\sc iso} surveys described in Elbaz et al. (1999) at 15 $\mu$m,
   and various analysis of {\sc iras} samples (see e.g. Guiderdoni et al. 1998
   at 60 $\mu$m).}  \label{figdiff}
\end{figure*}

Although the general agreement of our predicted counts with the
multi-wavelength data seems quite impressive, there are several
caveats.  At 15 microns, for example, one would say that we match the
integral counts fairly well (upper left panel of figure~\ref{figinf}). But
looking more closely, we cannot reproduce the change of slope seen in
the ISOCAM differential counts (fig.~\ref{figdiff}).  There are at least
a couple of reasons why this could happen.  First, the SEDs of the ISOCAM
galaxies are different in the mid-IR from the ones used here as a
template, which are based on {\sc iras} observations of the local universe.
Differential counts are very sensitive to the exact shape of the PAH
features, which are crudely modelled here. One could also imagine that
the discrepancy is partly due to the grain size distribution/chemical
composition evolution with redshift, as the redshift
distribution of ISOCAM galaxies has a median $z \approx$ 0.7. We plan
to investigate these issues in more detail but that is beyond the scope of
this paper.
Finally, because of the way interactions are
modelled, each early type galaxy undergoes a starburst after its host
halo has just collapsed, and it is not obvious that
ISOCAM sources (see figure~\ref{figinf}) in which the  vast majority
are LIRGs (not ULIRGs), are properly described by such a violent process.
Dynamical interactions (which are not
modeled in detail here), triggering multiple milder starbursts, with
time delays between them, might provide a more realistic description
of these sources and this is another issue we plan to investigate in the
future.

\subsection{Redshift distributions} 

\begin{figure*}
  \centerline{\psfig{file=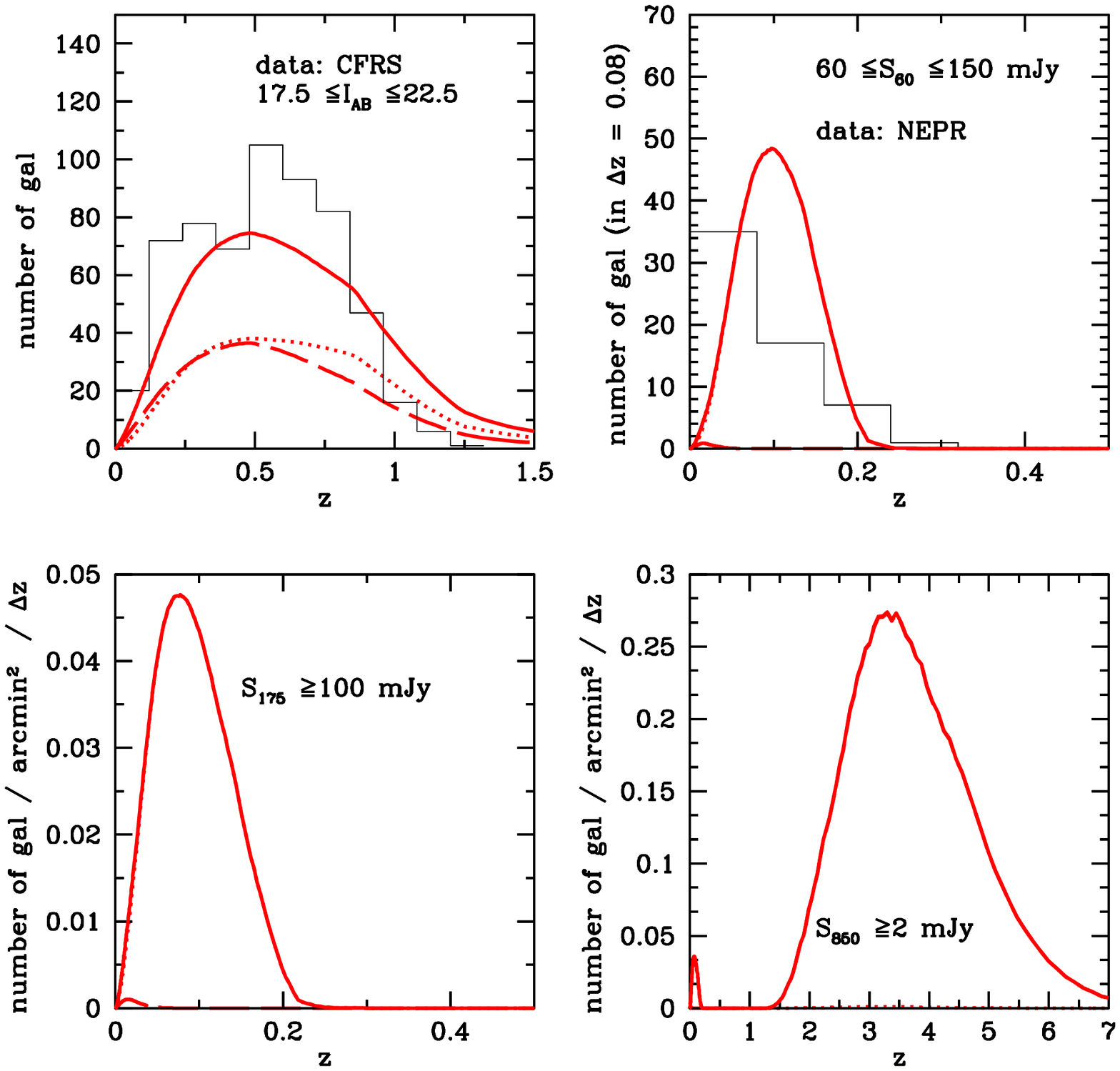,width=0.5\hsize}}
  \caption{Panchromatic redshift distributions of galaxies predicted
  by our analytic model.  Dotted curves represent late type galaxies
  (spirals); dashed: early type galaxies (ellipticals,
  S0s); and solid: the sum of both contributions.  Data in the I band
  (top left panel) is from Crampton et al. (1995) (Canada--France
  Redshift Survey), whereas at 60 microns, it comes from Ashby et
  al. (1996) (North Ecliptic Pole Region).  In both cases, the model
  curves have been renormalized to the total number of observed
  galaxies.}  \label{figred}
\end{figure*}

Another constraint on our models comes from  the redshift distributions.
Their shapes seem to quite nicely match the observations in the I band
with a mean redshift of the distribution of $\sim$ 0.6, (see top
left panel of figure~\ref{figred}). In the far--IR (60 microns), the agreement
with data gathered in the north ecliptic pole region (NEPR) is also
fairly convincing.  As predicted in Silk \& Devriendt (2000),
inclusion of the cosmological constant, $\Lambda$, has shifted the
near--IR and 60 micron peaks towards higher redshifts and produced a
high--redshift tail in the I band, bringing the models into closer
agreement with the data. Disks and early-types are found in comparable
proportion in the I-band, with a slight domination of disks for $z>0.3$ and up
to $z=1.5$.

 There are at least  a couple of reasons for
this behaviour. First, in this redshift range, spheroids are already
old, their star formation rates are very low and therefore their
I-band luminosity comes from an old and dim stellar population.
Secondly, spheroids in a massive starburst phase at these redshift
experience high dust absorption, which reduces 
significantly their luminosity even in the I band.
Going deeper in I magnitude leads to the selection
of higher redshift galaxies and we expect that the starburst
population to eventually dominate the distribution around $z \approx$ 2.

We draw the reader's attention to the dramatic change occuring in the
redshift distribution of sources when going from ISOPHOT to SCUBA,
{\em i.e.} from the far-IR, to the submm window (bottom panels of
figure~\ref{figred}). In the far-IR, the vast majority of sources is predicted to
lie at fairly low redshifts, whereas in the submm, there is a
characteristic double peaked distribution.
The first small peak at low redshift is
mainly due to disks, as can be seen on the lower right
panel of figure~\ref{figinf} ($\lambda_{eff}=850\ \mu$m).
Indeed, the Euclidian tail of the total counts (figure~\ref{figinf}) at fluxes
above $\sim 10^{-1}$ Jy on this figure is clearly due to
the domination of local bright disks (dotted line).
In contrast, one notes on the lower-right panel
of figure~\ref{figred}
an overwhelming domination of the (fainter) sources located at
high redshifts, as is also clear from figure~\ref{figinf}. 
Once again, the main driver for this effect is the negative k-correction
boosting observed fluxes in the submm wave-bands. Contrary to
Devriendt and Guiderdoni (2000) and Lagache et al. (2002), 
the second peak in the redshift
distribution of 175 microns sources (around $z=1$, see e.g.
Lagache et al. (2002)) is missing from
the model discussed here. This could either be a result of the
{\sc stardust} SEDs being too warm for LIRGs, or of a too rapid 
evolution in the number of LIRGs due to our simple modelling of 
galaxy interactions. Here too, triggering several starbursts
per galaxy with delays between them, as happens in realistic 
galaxy interactions, could alleviate the problem. 

\subsection{Diffuse background radiation} 

\begin{figure}
  \centerline{\psfig{file=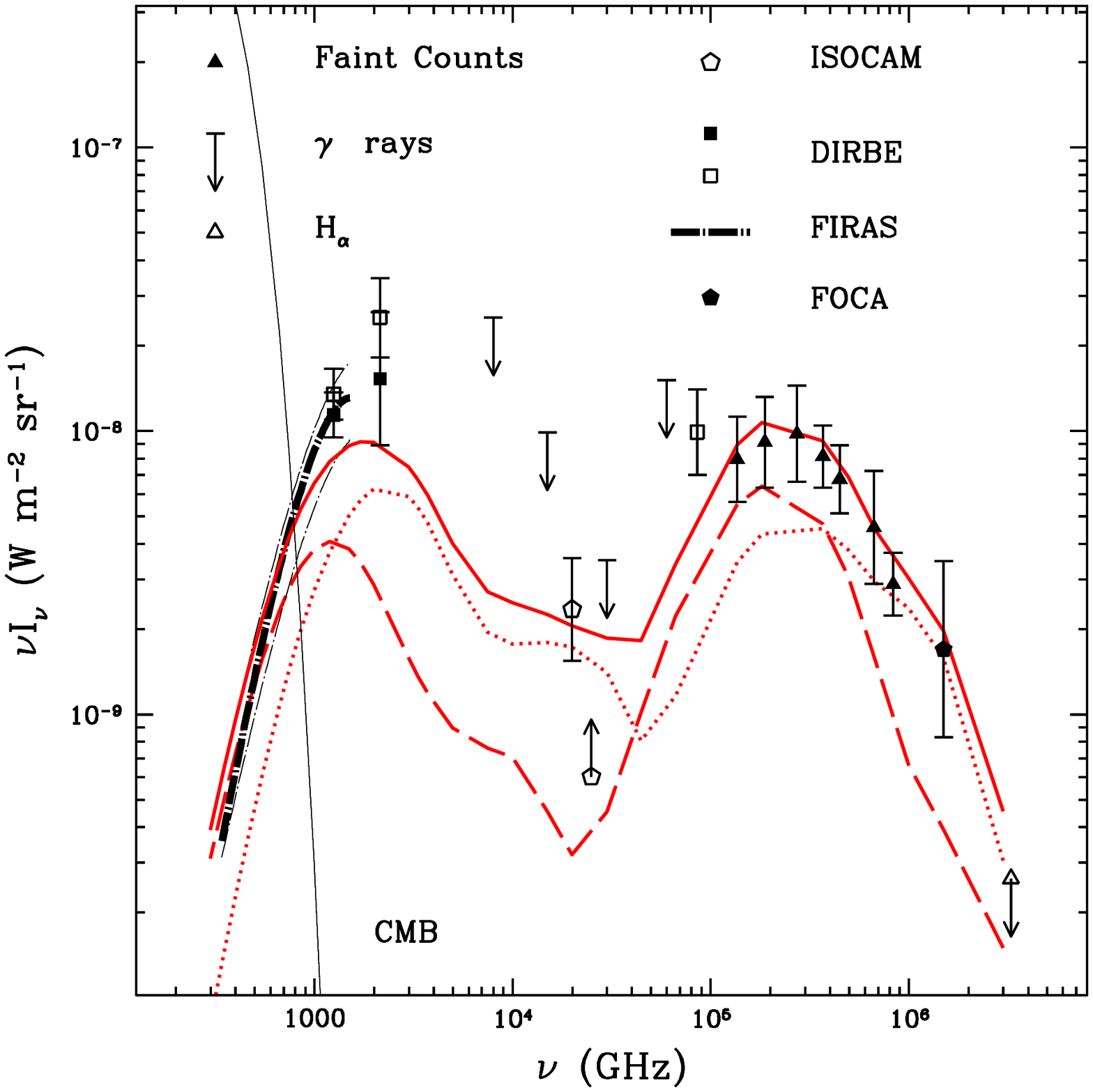,width=\hsize}} \caption{Diffuse
  background light emitted by star forming galaxies, from the UV to
  the submm.  Coding for the lines is dotted: late type galaxies
  (disks); dashed: early type galaxies (spheroids); and solid: sum of
  both contributions. The data sources are indicated on the figure.}
  \label{figback}
\end{figure}

The background light, as integrated from the multi-wavelength counts,
seems to closely match the detection by Puget et al. (1996) and Hauser
et al. (1998) in the far-IR/submm window (see
figure~\ref{figback}). This tells us that we are not grossly
overestimating or underestimating the faint counts, and that the
global galaxy luminosity budget from the UV to the submm is likely to
be correctly computed.  An interesting remark is that this plot
clearly shows that one has to go to the submm, around 1500 GHz (about
200 microns), to see the early-type galaxies dominate over the late
types.  This  is a factor 2 shorter in
wavelength than found for the SCDM model.  At any wavelength shorter
 than this  (except in the
near-IR between 2-5 microns), quiescent galaxies, forming up to
$\approx$ 10 M$_\odot$ of stars per year dominate the total light
emission (as well as the counts).  Again that was not the case with
the SCDM model where spirals dominated the whole background below 400
microns.
Finally, we point out that the model marginally underestimates 
the background measured by DIRBE. This discrepancy could stem from 
a fraction of the population of galaxies having  
slightly different spectral shapes in the far-IR than our model 
assumes they have, or having a redshift distribution different 
from the one we predict they have. 
Of course, as these effects are degenerate, in the sense that colder sources 
will emit more flux in the far-IR than hot ones, as will closer sources,
it is very difficult to decide which is the dominant without further guidance from 
the observations.

\subsection{Multi-wavelength comoving luminosity densities} 

\begin{figure}
  \centerline{\psfig{file=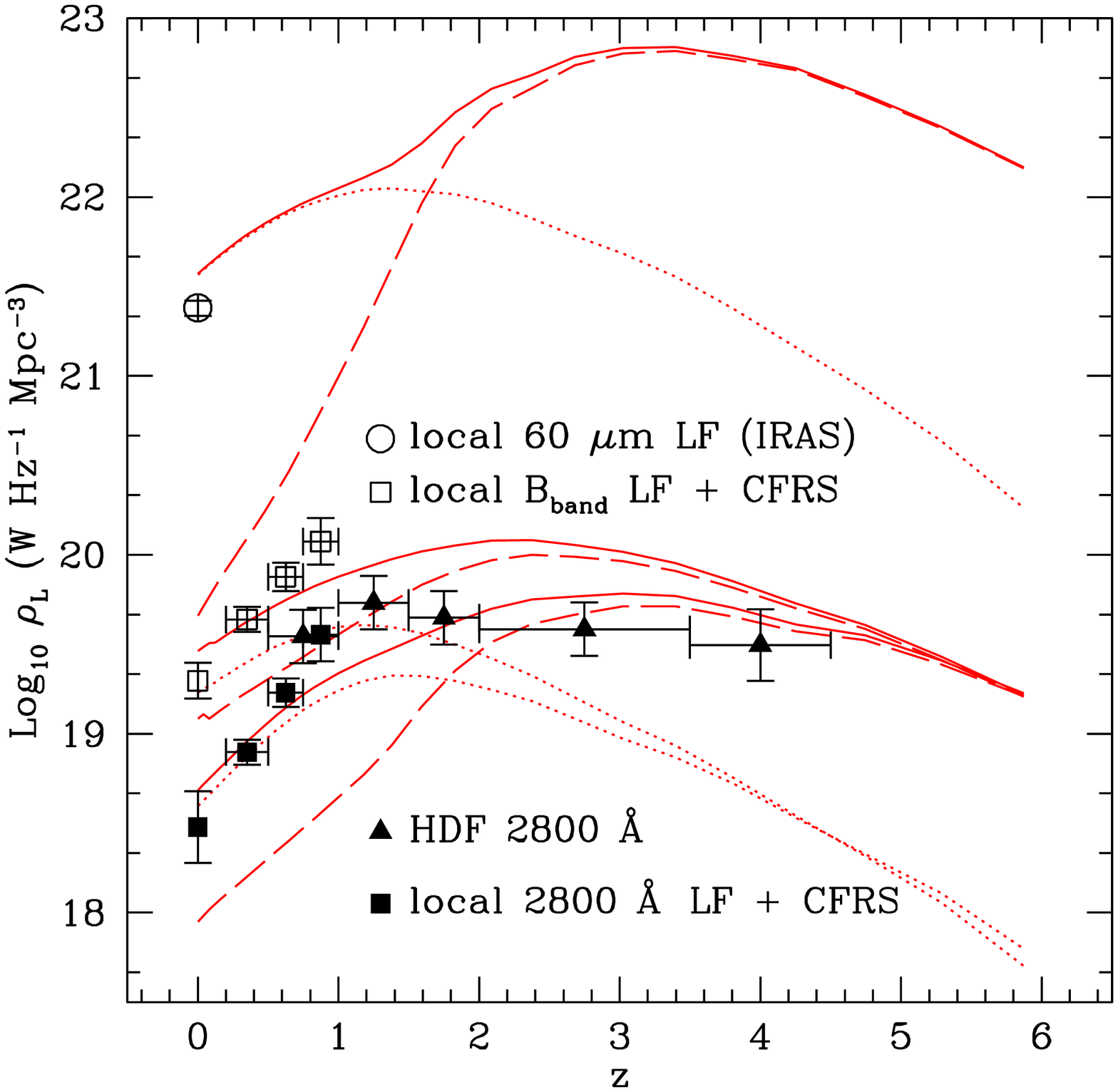,width=\hsize}} \caption{
Comoving luminosity densities as a function of redshift.
 As usual, coding for the lines is dotted: late type galaxies
  (disks); dashed: early type galaxies (spheroids); and solid: sum of
  both contributions. The three sets of curves represent predictions
of our model at 2800 \AA (bottom triplet), 4400 \AA (middle triplet) and
60 microns (top triplet).
  Data sources are indicated on the figure.}
  \label{figcolum}
\end{figure}

Figure~\ref{figcolum} shows the comoving luminosity densities
predicted by the model from the far-UV (bottom curves) to the far-IR
(top curves). It is clear from the figure that there already is a greater 
amount of evolution in the far-IR than in the UV-optical, and we know
from the counts (fig.~\ref{figinf}) that the submm must show still
more evolution than the far-IR. The agreement with the data is quite satisfactory
both at low and high redshift and for all wavelengths. This tells us
that the luminosity budget of our galaxies is relatively accurate.
Another striking feature is that the model predicts that the 
spheroid (whether in dusty starburst phase or not) contribution 
to the comoving luminosity density becomes approximately equal to that 
of late type galaxies around redshift 2 in the UV and far-IR 
and as early ar $z \sim$ 1 in the B-band. As a result, the peak of the 
comoving luminosity density shows a complex behavior as a function of
wavelength, starting around $z \sim$ 3 in the UV and falling to $z \sim$ 2
in the optical near-IR, before shifting again to higher redshifts
($z\sim$ 3) in the far-IR and submm. This reflects the fact that both
UV and far-IR luminosities are extremely sensitive to the
instantaneous SFR, whereas the optical and near-IR luminosities
are tainted by a non-negligible contribution coming
from an old stellar population.     

\subsection{Cosmic star formation history} 

\begin{figure}
\centerline{\psfig{file=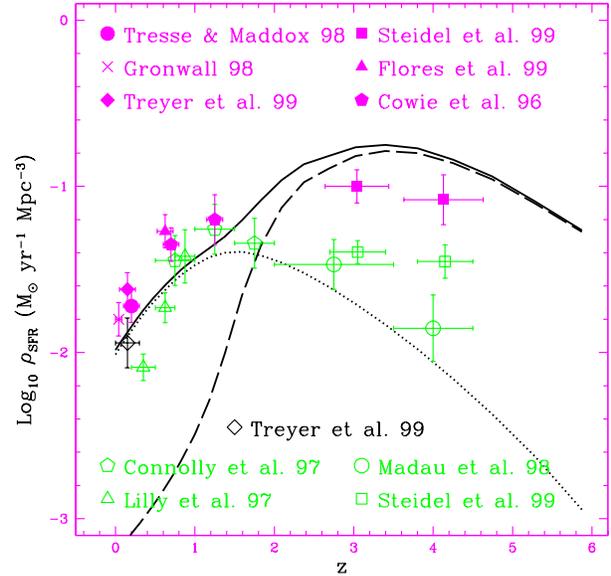,width=\hsize}}
\caption{Comoving star formation rate as a function of redshift.
  The different curves represent the contribution of late type
  galaxies (spirals: dotted line); early type galaxies (ellipticals,
  S0s: dashed line); and the sum of both contributions (solid line).
  Empty symbols give raw measurements of different teams (renormalized
  to match the cosmological model defined in table~\ref{cosmoparam}):
  circles (Madau, Pozetti \& Dickinson 1998); squares (Steidel et al.
  1999); triangles (Lilly et al. 1996) and pentagons (Connolly et al.
  1997).  Filled diamonds are data from Gronwall (1998); Tresse \&
  Maddox (1998); Treyer et al. (1998); Cowie, Songaila \& Barger
  (1999); Steidel et al. (1999) and are dust-corrected as described in
 Somerville et al. (2001).  The point at $z\sim 0.7$ is derived from far-IR
  ISOCAM sources (Flores et al. 1999). }

\label{figsfr}
\end{figure}

The predicted comoving star formation rate is compared to observations
on figure~\ref{figsfr}. Empty symbols are observed estimates
at various redshifts, uncorrected for dust extinction. Filled symbols
indicate dust corrected estimates following the prescription
of Somerville, Primack \& Faber (2001) based on a
luminosity dependant correction likely to be more realistic than
the traditional scaling by a constant factor (between 3 to 5)
at all $z$ (see their figure 9). The filled point at $z \sim 0.7$
is derived from infrared ISOCAM observations (Flores et al. 1999).

The comoving SFR is in fair
agreement with the measured one, and the reasonable amount of K-band
light produced (figure~\ref{figopt}) makes us confident that we do not
over-produce either stars or metals.  However note that at low
redshift, the agreement is only for non-dust-corrected data, so the
contribution of our spirals (which closely match the non corrected
Madau curve at {\em all} redshifts) might be
underestimated in our model.
From the perspective of metals only, the star formation
rate could be higher, provided that a fair fraction of the heavy
elements is ejected into the IGM. The model however predicts a peak in
star formation around redshift $\sim$ 3.5 (this is also true for the
SCDM case), sensibly higher than what is usually assumed,  current
data being compatible with a flat star formation rate at redshifts $z>$
1. A more detailed comparison with the SCDM cosmology used in Silk \&
Devriendt (2000) reveals that the comoving star formation rate increase
from $z = 0$ to $z \approx 3$ is steeper in the $\Lambda$CDM model,
due to the fact that fewer galaxies tend to form on average at low
redshifts in such a model. More
specifically, this increase in the steepness of the SFR results from:
\begin{itemize}
\item a similar high redshift SFR level in a $\Lambda$CDM and in a SCDM 
cosmology because the earlier formation of galaxies in a $\Lambda$CDM
model is compensated by a higher normalization of the power spectrum
in a SCDM model
\item  a lower contribution to the SFR history at late times in the
$\Lambda$CDM model due to the formation of a smaller number of galaxies
than in the SCDM model  
\end{itemize}
This explains why the discrepancy mainly shows up in the spiral population and
that, in contrast, the contribution of elliptical galaxies which form much earlier is
very similar in both cosmologies (Silk \& Devriendt 2000).  
We argue
that the amount of star formation occurring in dust-shrouded objects
cannot be much greater than our model predicts in order to avoid
overestimating the submm diffuse background (see
figure~\ref{figback}).  Therefore we emphasize that cosmic star
formation has to decrease at redshifts $z \ge$ 3.5.  Note that this
result is strengthened if dust-shrouded AGNs are a major contributor
to the submm emission, even though this does not appear to be the case
if X-ray characteristics are a reliable AGN monitor (e.g. Barger et al.
2001).

\section{Conclusions} 

We have implemented the model of BSS98 within the simple semi-analytic
model of DG00 in order to describe galaxy collisions on a physical
basis.  We find that such a combination:

\begin{itemize}

 \item naturally reproduces galaxy counts at various wavelengths and
 the diffuse background radiation.

 \item yields a fair match to existing redshift distributions ranging
 from the optical (CFRS) to far-IR ({\sc iras} at 60 microns). Furthermore,
 the model also predicts redshift distributions of sources at still
 larger wavelengths (ISOPHOT at 175 microns; SCUBA at 850 microns)

 \item predicts the contribution of morphological types to the
 previous quantities, with spheroids strongly dominating bright galaxy counts
 only in the sub-millimetre window (from 450 microns and longwards).

 \item predicts that the cosmic star formation rate history is
 dominated by spiral-like objects until redshift 2 where spheroids
 become the major contributors.

\end{itemize}
Our results qualitatively agree with the
model of the infrared universe
presented in Tan, Silk \& Balland (1999).
We also note  good agreement of the main conclusions of the present
work with the collisional starburst model of Somerville et
al. (2001). This agreement is examplified in the comparison
between our SFR plot (figure~\ref{figsfr} of the present paper)
and the SFR predicted by their collisional starburst
model (their figure 9).
It is satisfying that these two approaches give consistent
results as it supports the view that
galaxy-galaxy collisions in the high redshift universe
might have played a dominant role in triggering
star formation.

Note however that Somerville et al. fail to account for the submm
counts (Somerville private communication). They find too few highly obscured luminous galaxies.
Our model gives satisfactory agreement both in the restframe blue and FIR.
This is because we have the freedom, and indeed the
motivation from the model, to cumulate the effects of collisions. 
In contrast, semi-analytic hierarchical models form spheroids over a Hubble time at
the redshift at which galaxy mergers occur.
This may not be enough if the dust is dispersed by the ensuing starburst.
Introduction of an {\it ad hoc} delay between multiple starburst trigerred by different orbital stages 
during the pre-merger phase might solve this. Alternatively one might
appeal to additional sources of IR luminosity such as the one that might be associated
with AGN formation.
All indications however suggest that AGNs do not play a major role in
accounting for most of the observed submm sources, as evidenced from the
{\sc chandra} deep fields.
In our case we have
a physical model that works in the right direction
and yields a self-consistent panchromatic explanation of the observed
universe both locally and at high redshift.
We are aware that what we called spheroids at high redshifts but are in
the process of a violent burst of star formation will not have the morphologies
of a classical early type galaxy, as it will take some time for these objects to
relax. We defer a more detailed study of the evolution of relaxed galaxies 
to a companion paper, but note that it is likely that this fraction will be 
fairly small at $z \geq 3$ since there will be very little time elapsed
since the collision.

However, in spite of this issue, 
an important consequence of our model, is that the FIR counts and luminosity density are
expected to continue to rise with redshift to $z\sim 3.$ Hence we predict
that a future mission with adequate resolution to avoid source confusion,
 which sadly may not be the
case for {\sc astrof} or {\sc sirtf},
will provide an important test of the galaxy collision model.

\section*{Acknowledgments}

We thank Guilaine Lagache, Herv\'e Dole and Jean-Loup Puget for
providing us with their results before publication. We also acknowledge 
a constructive referee report from which the presentation of the 
model has benefited a lot. J.D. is supported by the Leverhulme trust.


\end{document}